\documentclass[11pt,]{article}
\pdfoutput=1
\usepackage{authblk}
\usepackage{fullpage}
\usepackage{amssymb,amsmath}
\usepackage[utf8]{inputenc}
\usepackage[T1]{fontenc}
\usepackage{siunitx}
\usepackage{float}
\usepackage[square,numbers]{natbib}

\usepackage{setspace}
\usepackage{graphicx}

\makeatletter
\def\ScaleWidthIfNeeded{%
 \ifdim\Gin@nat@width>\linewidth
    \linewidth
  \else
    \Gin@nat@width
  \fi
}
\def\ScaleHeightIfNeeded{%
  \ifdim\Gin@nat@height>0.9\textheight
    0.9\textheight
  \else
    \Gin@nat@width
  \fi
}
\makeatother
\setkeys{Gin}{width=\ScaleWidthIfNeeded,height=\ScaleHeightIfNeeded,keepaspectratio}%

\title{Microscale velocity-dependent unbinding generates a macroscale performance-efficiency tradeoff in actomyosin systems}
\author[a]{Jake McGrath}
\author[a,b]{Brian Kent}
\author[a]{Colin L. Johnson}
\author[a]{José Alvarado*}
\affil[a]{Center for Nonlinear Dynamics, Department of Physics, University of Texas at Austin, 2515 Speedway, Austin, Texas 78712, USA.}
\affil[b]{Theory Group, Weinberg Institute, Department of Physics, University of Texas, 2515 Speedway, Austin, Texas 78712, USA.}
\date{\today}

\begin{document}

\maketitle
\vspace{-2em}
\textsuperscript{*}To whom correspondence should be addressed. E-mail: alv@chaos.utexas.edu.
\begin{abstract}
\setstretch{1.5}
Myosin motors are fundamental biological actuators, powering diverse mechanical tasks in eukaryotic cells via ATP hydrolysis. Recent work revealed that myosin’s velocity-dependent detachment rate can bridge actomyosin dynamics to macroscale Hill muscle predictions. However, the influence of this microscale unbinding, which we characterize by a dimensionless parameter $\alpha$, on macroscale energetic flows—such as power consumption, output and efficiency—remains elusive. Here we develop an analytical model of myosin dynamics that relates unbinding rates $\alpha$ to energetics. Our model agrees with published \textit{in-vivo} muscle data and, furthermore, uncovers a performance-efficiency tradeoff governed by $\alpha$. To experimentally validate the tradeoff, we build HillBot, a robophysical model of Hill’s muscle that mimics nonlinearity. Through HillBot, we decouple $\alpha$’s concurrent effect on performance and efficiency, demonstrating that nonlinearity drives efficiency. We compile 136 published measurements of $\alpha$ in muscle and myoblasts to reveal a distribution centered at $\alpha^* = 3.85 \pm 2.32$. Synthesizing data from our model and HillBot, we quantitatively show that $\alpha^*$ corresponds to a class of generalist actuators that are both relatively powerful and efficient, suggesting that the performance-efficiency tradeoff underpins the prevalence of $\alpha^*$ in nature. We leverage these insights and propose a nonlinear variable-impedance protocol to shift along a performance-efficiency axis in robotic applications.
\end{abstract}
\newpage

\section{Introduction}
Myosin motors are ubiquitous, fundamental molecular machines responsible for powering many biological processes including intracellular transport, cytokinesis, and muscle contraction. These motors operate at submicron scales, utilizing chemical energy from ATP hydrolysis to generate mechanical force, yet their force production is vital across scales from cellular functions to organismal-level activities. By coordinating in vast ensembles, myosin motors drive macroscopic muscle dynamics, enabling a diverse set of mechanical tasks such as locomotion (walking, swimming, flight), internal regulation (stability, balance, fluid transport), and athletics (endurance, high-intensity). Myosin coordination on the organismal-scale is adaptable, either on long time scales due to selection pressures \cite{oneill_chimpanzee_2017,boyle_muscles_2020} or on short time scales due to training regimen \cite{haugen_sprint_2019,plotkin_muscle_2021,wilson_effects_2012}. In order to power cellular and organismal dynamics, myosin motors convert a stored energy source, in the form of adenosine triphosphate (ATP), into useful mechanical work. How myosin’s energy transduction mechanisms, which are ultimately microscopic (nm) in origin, manifest on energetic flows at cellular ($\mu$m) and organismal scales (m) remains a formidable research challenge.

A multitude of mechanisms determine how microscopic activity transduces chemical free-energy consumption to macroscopic mechanical work output. Various biochemical and structural factors—such as muscle-specific protein isoforms and distinct metabolic pathways to supply ATP \cite{syrovy_isoforms_1987}, and microstructural features like post-stimulation transients \cite{hill_abrupt_1997}, variable thick-thin-filament overlap \cite{gordon_variation_1966}, filament length polydispersity \cite{edman_sarcomere_1987}, length-dependent calcium concentration \cite{stephenson_length_1984}, advection of substrates \cite{cass_mechanism_2021}, and sensitivity to small changes in lattice spacing \cite{tune_nanometer-scale_2020}—modulate how microscopic myosin kinetics translates into macroscopic mechanical output. These microscopic influences shape key muscle characteristics observed at larger scales \cite{zajac_muscle_1989}, including the force-length curve \cite{maganaris_forcelength_2001}, force response to stimuli \cite{sanderson_electrical_1895}, and work loops \cite{josephson_mechanical_1985}. However, ultimately, it is the chemomechanical activity of actomyosin cross-bridges, where myosin-mediated ATP hydrolysis couples to the relative sliding of actin, that drives actuation in biology.

A distinctive feature of the myosin power stroke, relative to other biochemical reactions (such as typical enzymatic reactions or glycolysis), is myosin’s velocity-dependent off-rate \cite{piazzesi_skeletal_2007} wherein faster contraction rates cause quicker unbinding. Recent work has shown that this velocity-dependent unbinding aligns myosin kinetics with Hill’s historical muscle model, connecting microscopic activity to macroscopic force-velocity characteristics through a nonlinearity parameter $\alpha$ \cite{seow_hills_2013}. Here, the nonlinearity parameter, $\alpha$, functions as a coarse-grained descriptor translating microscopic myosin activity rates into macroscopic force-velocity curvature, as characterized by Hill’s model \cite{hill_heat_1938}.

Seminal work by Hill \cite{hill_heat_1938} experimentally measured muscle's force-velocity (FV) curve, and found a nonlinear, hyperbolic relation. Although Hill’s experiments were performed under quick-release from isometric conditions \cite{hill_heat_1938}, and Hill’s model is not precise near stall \cite{edman_double-hyperbolic_1988,lou_high-force_1993}, or under eccentric contractions \cite{krylow_dynamic_1997}, it remains a reasonable approximation under quasistatic conditions and is frequently used in the region of maximum power for many kinds of motions \cite{seow_hills_2013,cadova_comparative_2014}. In normalized form, Hill's force-velocity relation is shown in Eq. 1
\begin{equation}
    \label{eq:eq1}
    F = \frac{{1 - V}}{{1 + \alpha V}}
\end{equation}
where the muscle's tension $f$ and contraction rate $v$ are normalized by isometric tension $f_m$ and no-load speed $v_m$ to yield dimensionless quantities \( F=f/f_m,V=v/v_m \in [0,1] \) (see SI Table 1 for a full list of variable definitions) and the non-negative, dimensionless parameter \( \alpha \geq 0 \) governs nonlinearity (Fig. 1b).

Within the framework of Hill's model, empirical findings have demonstrated how a concave force-velocity relation shapes energetics. Studies have correlated efficiency with higher nonlinearities and have shown that heat generation in contracting muscle fibers depends on $\alpha$ \cite{hill_efficiency_1964,woledge_energetics_1968}. Early studies suggested that more curved FV relations give fewer negatively strained cross-bridges \cite{woledge_energetics_1968}, which would tend to unbind once a certain conformation was reached \cite{eisenberg_relation_1980}, with X-ray diffraction studies confirming this mechanism \cite{piazzesi_skeletal_2007}.

Although empirical connections between concavity and energetics in the context of Hill's model have been made, analytical connections between the microscale velocity-dependent unbinding of myosin and macroscale energetic flows—from free-energy consumption to mechanical power output—remain elusive.

Here, we develop an analytical 2-state model of actomyosin dynamics that relates molecular myosin unbinding rates to emergent macroscale energetics. We demonstrate that our model agrees with published \textit{in-vivo} muscle data (both in power outputs and energetic efficiencies). Furthermore, the model uncovers a performance-efficiency tradeoff governed by myosin's unbinding rate $\alpha$. To experimentally investigate the tradeoff, we build HillBot, a robophysical model of Hill’s muscle, as experimentally measuring energy consumption and muscle mechanics \textit{in-vivo} is difficult and often requires estimation \cite{erdemir_model-based_2007,buchanan_neuromusculoskeletal_2004,buchanan_estimation_2005,davy_dynamic_1987}. Moreover, it is difficult to experimentally test a series of muscles with systematically varying nonlinearities $\alpha$ while keeping all other properties constant—e.g., the muscle’s maximum force, contraction-rate, ratio of fast- to slow- twitch fibers, training-level, fatigue, cross-sectional area, length, etc. Through HillBot, we decouple $\alpha$’s concurrent effect on performance and efficiency, demonstrating that concavity in muscle's force-velocity relation drives efficiency at the sacrifice of power output. Finally, we compile 136 published measurements of $\alpha$ in muscle and myoblasts to reveal a distribution centered at $\alpha^* = 3.85 \pm 2.32$. Synthesizing data from our model and HillBot, we quantitatively show that $\alpha^*$ coincides with a minimum of a cost function representing the performance-efficiency tradeoff. Finally, leveraging these insights, we discuss a nonlinear variable-impedance protocol to shift along a performance-efficiency axis in robotic applications.

\section{Results}
\subsection{Cross-bridge dynamics control energetic flows in muscle}
We build a simple two-state model of muscle contraction following \cite{piazzesi_skeletal_2007, seow_hills_2013} to study how the nonlinear, velocity-dependent unbinding rate of cross-bridges affects energy consumption and mechanical power generation.

The model starts following Ref. \cite{seow_hills_2013}, which connects microscopic actin-myosin dynamics to macroscopic Hill dynamics. We start by assuming a simple 2-state model of actin-myosin contraction \cite{huxley_6_1957} where some proportion of myosin are attached to actin $A$ and some are detached $D$ (Fig. 1a). The sum of detached and attached myosin proportions must be 1, $A+D=1$. We assume the rate of myosin attachment $k_a$ to be constant and velocity-independent. We further assume the rate of myosin detachment $k_d(V)=K_DV+k_0$ to be velocity-dependent, with a finite detachment rate $k_0$ at zero velocity (i.e. the isometric detachment rate) \cite{piazzesi_skeletal_2007}. We write down the rate at which the fraction of attached myosin change as
\begin{equation}
    \dot{A} = k_aD - k_dA
\end{equation}
Non-dimensionalizing the above equation yields
\begin{equation}
    \dot{A} = 1 - A(1 + \alpha V + \gamma)
\end{equation}
where time is rescaled by myosin's attachment rate $\hat{t}=tk_a$ and the nonlinearity is defined as $\alpha=K_D/k_a$ \cite{seow_hills_2013} and $\gamma = k_0/k_a$. Experimental data of the anterior tibialis muscle in frog \cite{piazzesi_skeletal_2007} demonstrate $\gamma \approx 1$.

Next, in order to couple myosin binding dynamics to actuation, we let this simple 2-state model of actin-myosin contract against some mass $m$
\begin{equation}
    m\dot{v} = f_0 A (1 - V)
\end{equation}
where the contracting muscle's force is given by the product of the fraction of attached, force-producing myosin $A$, the stall force $f_0$ of an individual motor, and the linear force-velocity relation of a single myosin motor \cite{piazzesi_skeletal_2007}. We again rescale time $\hat{t}=tk_a$ and normalize acceleration $\dot v$ by $v_mk_a$ to arrive at
\begin{equation}
    \dot{V} = \beta A (1 - V)
\end{equation}
where $\beta=f_0/mv_0k_a$ is a dimensionless quantity relating muscle actuation properties to the inertia of the attached mass. When attached to a mass $m = 10^{-1}  \ \mathrm{kg}$, a rough order of magnitude estimate suggests $\beta \approx 1$ given that $k_a = 10 \ \mathrm{s^{-1}}$ \cite{piazzesi_skeletal_2007}, $f_0 = 78 \ \mathrm{mN}$ and $v_0 = 27 \ \mathrm{mm \ s^{-1}}$ in the soleus of a mouse \cite{houdijk_evaluation_2006}.

Although our model includes the fact that individual myosin motors have a linear FV-curve \cite{piazzesi_skeletal_2007}, the velocity-dependent myosin detachment rate $K_d$ introduces a nonlinear coupling between the equations of motion for $\dot A$ and $\dot V$. The dimensionless nonlinearity parameter $\alpha = K_D / k_a$ directly relates to the curvature parameter $\alpha = f_m/a$ of Hill’s original model, as demonstrated in steady state in Ref. \cite{seow_hills_2013}. Here, we confirm that the transient dynamics of our model can closely follow the hyperbolic relation originally described by Hill \cite{hill_heat_1938}  (Fig. 1b, 1c). Recognizing $F=\beta^{-1}\dot V$ and taking $\gamma \rightarrow 0$, Eq. 2 in steady state yields exactly Hill's equation (Eq. 1). We have thus shown that our simplified system of equations is sufficient to describe how the nonlinear coupling between myosin unbinding dynamics and actuation, expressed via $\alpha$, captures the kinematics in force-velocity space described by Hill’s equation.

Next, we systematically assess how nonlinearity controls macroscopic energetic flows. For this, we include an additional equation that describes how $\alpha$ affects energy consumption in muscle, both directly by ATPase from the power stroke as well as from maintenance heat.

Myosin hydrolyzes ATP, resulting in the release of Gibbs free energy $g_0$. As the total number of attaching myosin $N k_a (1 - A)$ transition to the attached state, the total rate of Gibbs free-energy consumption $\dot{g}$ is given by
\begin{equation}
    \dot{g} = g_0 N k_a (1 - A)
\end{equation}
In addition to utilizing Gibbs free energy for contraction, muscle also produces maintenance heat for homeostatic regulation, $\dot{h_m}$. Experiments have shown that the rate of maintenance heat production is approximately $\dot{h_m}=ab$ where $a,b$ are constants in Hill's muscle model \cite{hill_efficiency_1964}. Therefore, the total rate of energy liberation in the model is given as
\begin{equation}
\begin{split}
    \dot{e} &= \dot{g} + \dot{h_m} \\
    \dot{e} &= g_0 N k_a (1 - A) + ab
\end{split}
\end{equation}
Which, after rescaling by a characteristic system power $f_m v_m$, and noting that $av_m=bf_m$ for $\alpha_{\rm Hill}=f_m/a$ in Hill's original model, we arrive at
\begin{equation}
    \dot{E} = \zeta (1 - A) + \frac{1}{\alpha^2}
\end{equation}
where $\zeta = N k_a g_0 / f_m v_m$ relates the rate of Gibbs free-energy release due to myosin ATPase activity to the rate of mechanical work output. Recognizing that $f_m v_m \approx N f_0 v_0$, with $f_0$ and $v_0$ the maximum force and velocity of an individual myosin motor, we have $\zeta \approx \frac{k_a g_0}{f_0 v_0} = 1/3$, given $g_0 = 100 \ \mathrm{zJ}$ \cite{milo_cell_2015}, $k_a = 10 \ \mathrm{s^{-1}}$, $f_0 = 6 \ \mathrm{pN}$ and $v_0 = 10 \ \mathrm{\mu m \, s^{-1}}$ for single motors \cite{piazzesi_skeletal_2007}. The rate of energy release in our 2-state model is in agreement with prior phenomenological models which found that the heat generated during contraction is proportional to the sum of maintenance heat and the rate of shortening \cite{fitzhugh_model_1977, umberger_model_2003}.

Equations 3, 5, and 8 represent our simplified model that describes how myosin’s nonlinear unbinding affects energetics. We integrate these equations forward in time, setting $\beta=\gamma=\zeta=1$ but allowing the unbinding rate $\alpha$ to vary. Fig. 1d shows a time series of the fraction of attached myosin motors $A$ and the normalized force and velocity for $\alpha=2$. Increasing the nonlinearity decreases the fraction of attached myosin (Fig. 1e), reducing power output (Fig. 1c). 
\begin{figure}[H]
    \centering
    \includegraphics[width=0.75\textwidth]{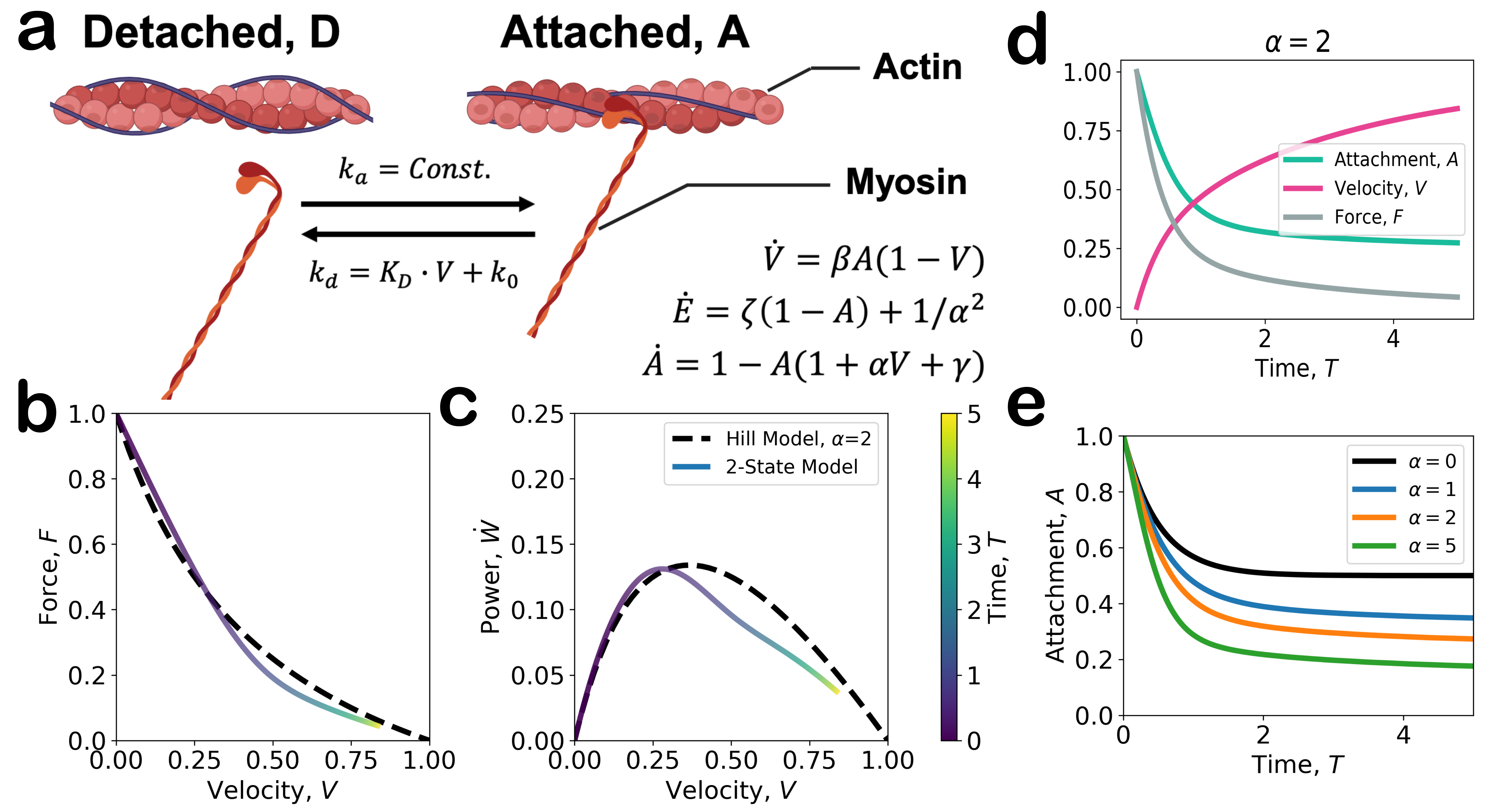}
    \caption{A nonlinear model accounting for 2-state myosin binding dynamics and actuation of a mass. a) Schematic of the 2-state model of myosin binding dynamics in muscle. Myosin cycles between detached and attached states. b), c) Time series of actuation, represented in force-velocity (b) and power-velocity (c) spaces compared with the predictions of Hill's model in Eq. 1 (black-dashed line). Color corresponds to $T$ (color bar, right). d) Time series of model attachment (teal line), velocity (magenta line), and force (gray line) for $\alpha=2$, with $T=k_at$ dimensionless time. e) Time series of myosin attachment for $\alpha = 0,1,2,5$ (black, blue, orange, and green lines, respectively); a larger fraction of myosin unbind with increasing nonlinearities.}
    \label{fig:f1}
\end{figure}

To characterize the total power of the actuator independent of actuation time or task, we introduce the power characteristic $PC$. The $PC$ is a scalar that quantifies an actuator’s ability to output power at all contraction rates, and we define $PC$ as the integral of an actuator’s normalized force-velocity curve. Using Hill’s equation (Eq. 1), we have:
\begin{equation}
    \label{eq:e2}
    PC := \int_{0}^{1} F \cdot \, dV = \frac{(1+\alpha) \cdot \ln(1+\alpha) - \alpha}{\alpha^2}
\end{equation}
where $PC$ $\to 1/2$ in the limit of linear actuation $\alpha \to 0$. We show that the total mechanical work delivered by our 2-state model in a fixed amount of time increases with $PC$ (SI Fig. 1d, Fig. 2a, solid lines) since more powerful actuators have higher work capacities.

In addition to the model's mechanics $PC$, we calculate the model's energetic efficiency at each time point as a ratio of instantaneous power generation to the rate of energy release, $\eta_\text{inst}=\frac{\dot{W}}{\dot{E}}$ (Fig. 2b). Moreover, to further validate the energetics of our model, we identify $N = 26$ measurements across 11 studies, where efficiency and $\alpha$ were concurrently measured over nine distinct muscle groups \cite{woledge_energetics_1968, hill_efficiency_1964,barclay_mechanical_1996,barclay_energetics_1993,houdijk_evaluation_2006,hill_mechanical_1939,gilbert_tension_1978,gilbert_effect_1986,buschman_mechanical_1997,curtin_efficiency_1991,rall_energetics_1973} (Fig. 2d, SI Table 4). Although some studies define efficiency as the ratio of work $W$ to energy liberated $E$ (or, work to the sum of work and heat) \cite{woledge_energetics_1968, barclay_energetics_1993,barclay_mechanical_1996,houdijk_evaluation_2006}, here we consider the maximum instantaneous efficiency during actuation transients. Because we interpolate the data across different studies, the work done by each muscle sample is not standardized. Reporting total efficiency (work to the sum of work and heat) may be difficult to compare as the muscle samples may have spent different amounts of time in steady-state. Therefore, for the following analysis, we define $\eta = \max_{V \in [0,1]} \frac{\dot{W}}{\dot{E}}$ as this metric standardizes the efficiency data for different studies. Additionally, for each of the $N=26$ efficiency and $\alpha$ measurements, we calculate $PC$ for each muscle sample by Eq. 9 (Fig. 2c). We find that, by visual inspection, our model agrees well with the $PC$ and $\eta$ results of our meta-analysis (Fig. 2c, 2d), demonstrating that coupling cross-bridge dynamics to transient dynamics of an accelerated object suffices to account for nonlinear energetic flows (inputs to outputs) in muscle.

For increasing nonlinearity $\alpha$, we demonstrate that a tradeoff between $PC$ and $\eta$ emerges (Fig. 2c, 2d): as curvature in FV-space increases with $\alpha$, the work capacity $PC$ decreases yet energetic efficiency $\eta$ increases. In the linear limit $\alpha \rightarrow 0$, the ratio of myosin detachment rates to attachment rates decreases, leaving the myosin mostly attached (Fig. 1e) and generating force. In this limit, we recover the highest PC (Fig. 2c). However, with relatively high myosin attachment rates, ATP consumption and heat generation increases (Fig. 2a). Increasing nonlinearity causes more cross-bridges to unbind (Fig. 1e). Although this reduces power (Fig. 2c), it increases efficiency (Fig. 2d) by reducing ATP consumption and heat generation.
\begin{figure}[H]
    \centering
    \includegraphics[width=0.6\textwidth]{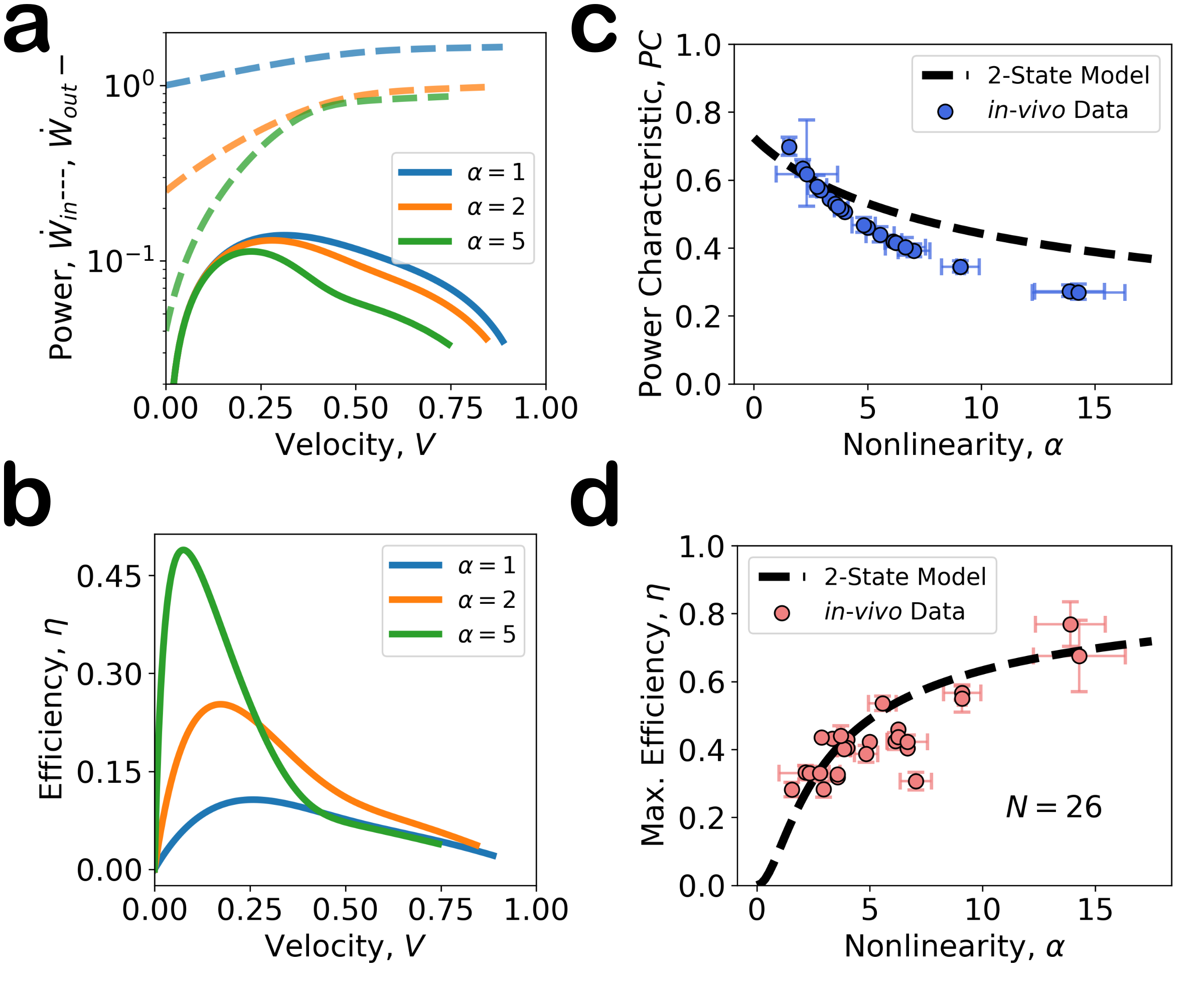}
    \caption{A performance-efficiency tradeoff emerges. a) As the rate of unbinding ($\alpha$) increases, mechanical power output (solid lines) and power consumption (dashed lines) decreases, however, because power consumption drops at a faster rate than power output, b) we find increasing efficiency with the nonlinearity. c), d) Performance-efficiency tradeoff emerges in 2-state model (black-dash lines) and \textit{in-vivo} data taken from the literature (scatter points); as the unbinding rate $\alpha$ increases, the actomyosin system is less powerful (lower $PC$) yet it gains higher instantaneous efficiency $\eta$ values.}
    \label{fig:f2}
\end{figure}
These results quantitatively elucidate the molecular basis for a tradeoff between efficiency, $\eta$, and performance, $PC$, in actin-myosin systems while demonstrating strong agreement with \textit{in-vivo} muscle data taken from the literature. In addition to our analytical 2-state model, to experimentally assess how nonlinear unbinding $\alpha = K_D / k_a$ governs a tradeoff in performance and efficiency keeping all other muscle parameters constant, we develop HillBot, a robophysical model of the Hill muscle.
\subsection{HillBot elucidates a performance-efficiency tradeoff}
In order to quantitatively assess how the nonlinearity determines power and efficiency in experiment, we extend our earlier work \cite{mcgrath_hill-type_2022}. We develop a robophysical muscle model called HillBot (Fig. 3a, 3b, 3c) consisting of a DC motor and an Arduino microcontroller that utilizes proportional-integral-derivative (PID) feedback control (SI Fig. 1a) to mimic Hill’s nonlinear force-velocity profile for a user-specified nonlinearity $\alpha$ (Fig. 3d, 3e, 3f).

We employ HillBot as it is difficult to experimentally test a series of muscles with systematically varying values of the velocity-dependence $K_D$ on the myosin unbinding rate (i.e. $\alpha = K_D/k_a$) while keeping all other properties constant, e.g., the muscle’s maximum force, contraction-rate, ratio of fast- to slow- twitch fibers, training-level, fatigue, cross-sectional area, length, among other muscle properties. In addition, accessing muscle state variables (the rate of energy consumption $\dot E$ and power output $\dot W$, along with others) in real time is difficult to measure in living muscle and often requires estimation \cite{erdemir_model-based_2007,buchanan_neuromusculoskeletal_2004,buchanan_estimation_2005,davy_dynamic_1987}. Through HillBot, we can experimentally measure energetics in real time while keeping all muscle properties constant for a range of $\alpha$, allowing us to have experimental control in monitoring $\alpha$’s concurrent effects on efficiency and performance which is otherwise difficult with \textit{in-vivo} tissue.

We construct HillBot's differential equations with Newton’s and Kirchhoff’s second laws to describe its electo-mechanical behavior (Fig. 3b). To couple these equations, we use the fact that the back EMF is proportional to the motor’s output angular velocity by a constant $k_1$ and the current draw is proportional to the motor’s output torque by a constant $k_2$. To achieve Hill-type actuation, the PID minimizes the error function in Eq. 13 (defined as the difference between the predicted Hill muscle force in Eq. 1 and the measure output force of HillBot) by dynamically reducing the voltage in Eq. 12 (Fig. 3g), similar to the myosin unbinding from earlier.
\begin{equation}
    \dot{\omega}(t) = \frac{1}{J} \cdot (k_2i - b\omega)
\end{equation}
\begin{equation}
    \dot{i}(t) = \frac{1}{L} \cdot (\mathcal{E} - iR - k_1\omega)
\end{equation}
\begin{equation}
    \mathcal{E}(t) = k_p e(t) + k_i \int_{0}^{t} e(\tau) \, d\tau + k_d \frac{de(t)}{dt}
\end{equation}
\begin{equation}
    e(t) = 12 \cdot \left( \frac{1-V}{1+\alpha V} - F \right)
\end{equation}

The applied voltage $\mathcal{E} \in [0,12]$ (in Volts) is determined by the measured current draw $i$ (in Amps) and angular velocity $\omega$ (in Rads/s) (which are proportional to the normalized force $F$ and velocity $V$ of the motor; see Methods). The PID gains ($k_p,k_i,k_d$) are set by the user beforehand and are thus known constants. The resistance, inductance, angular velocity-back EMF constant, torque-current constant, moment of inertia, and damping constants (written as $\Pi=\{R,L,k_1,k_2,J,b\}$) are unknowns, which we determine through fitting (see Methods, SI Fig. 1c).

For a range of $\alpha$, we integrate Eq. 10-13 forward in time. Also, varying $\alpha$ across $N=169$ trials, HillBot lifts a small mass $m$ and we record HillBot’s angular displacement and current draw every $\sim 10~\mathrm{ms}$ until the system reaches steady state ($T = 0.68~\mathrm{s}$). We extract the system's state, expressed in terms of the force $f$ applied and the velocity $v$ with which the mass rises (Fig. 3d). We verify that the model and HillBot are consistent with Hill’s equation (Eq. 1; Fig. 3e, 3f, SI Fig. 1b), and find strong agreement, demonstrating that our system accurately mimics the mechanics of Hill's model by dynamically "unbinding" the voltage (Fig. 3g).

During a lifting task, we quantify HillBot's total energy consumption $E$ by integrating a time series of electrical power consumption $\dot E = \mathcal{E} \cdot i$ up to steady state time $T$. Similarly, we integrate a time series of the actuator's mechanical power $\dot{W} = f \cdot v$ to calculate its total work output $W$. We then define HillBot's efficiency during actuation as the ratio of work to energy consumption, $\eta = W/E$. Furthermore, we quantify HillBot's mechanics through the power characteristic $PC$ in Eq. 9 as before.

We observe that $PC$ decreases as $\alpha$ increases (Fig. 3i). This is consistent with the observation that increasing $\alpha$ increases the curvature of the actuator's FV-curve (Fig. 3e) and hence suppresses power outputs (Fig. 3f). We also note that $\eta$ increases with $\alpha$ (Fig. 3h); this result aligns with our previous findings \cite{mcgrath_hill-type_2022} where we varied $\alpha$ while holding HillBot's work output constant. At the onset of each trial, the controller described in Eq. 12 applies a high initial voltage in response to a large initial error in Eq. 13. As the motor begins to rotate, to achieve a response consistent with the functional form of Hill's model (Eq. 1), the controller subsequently decreases voltage to moderate force output$-$higher nonlinearities require larger "unbinding" of voltage (Fig. 3g). This voltage reduction (as overestimated by the model in Fig. 3d), coupled with the motor’s inertia sustaining its rotation, facilitates efficiency gains with the nonlinearity, as demonstrated in Fig. 3h.

Increasing $\eta$ and decreasing $PC$ as the unbinding rate $\alpha$ increases mirrors the main results of the 2-state model from earlier (Fig. 2c, 2d): through HillBot, we again find a performance-efficiency tradeoff (Fig. 3h, 3i) parameterized by the nonlinearity $\alpha$. Although HillBot and muscle consume energy by different mechanisms, we show that dynamically "unbinding" to generate Hill-type dynamics (whether it be through voltage (Fig. 3g) or cross-bridges (Fig. 1e)) generates energetic efficiency at the sacrifice of power output.

\begin{figure}[H]
    \centering
    \includegraphics[width=0.7\textwidth]{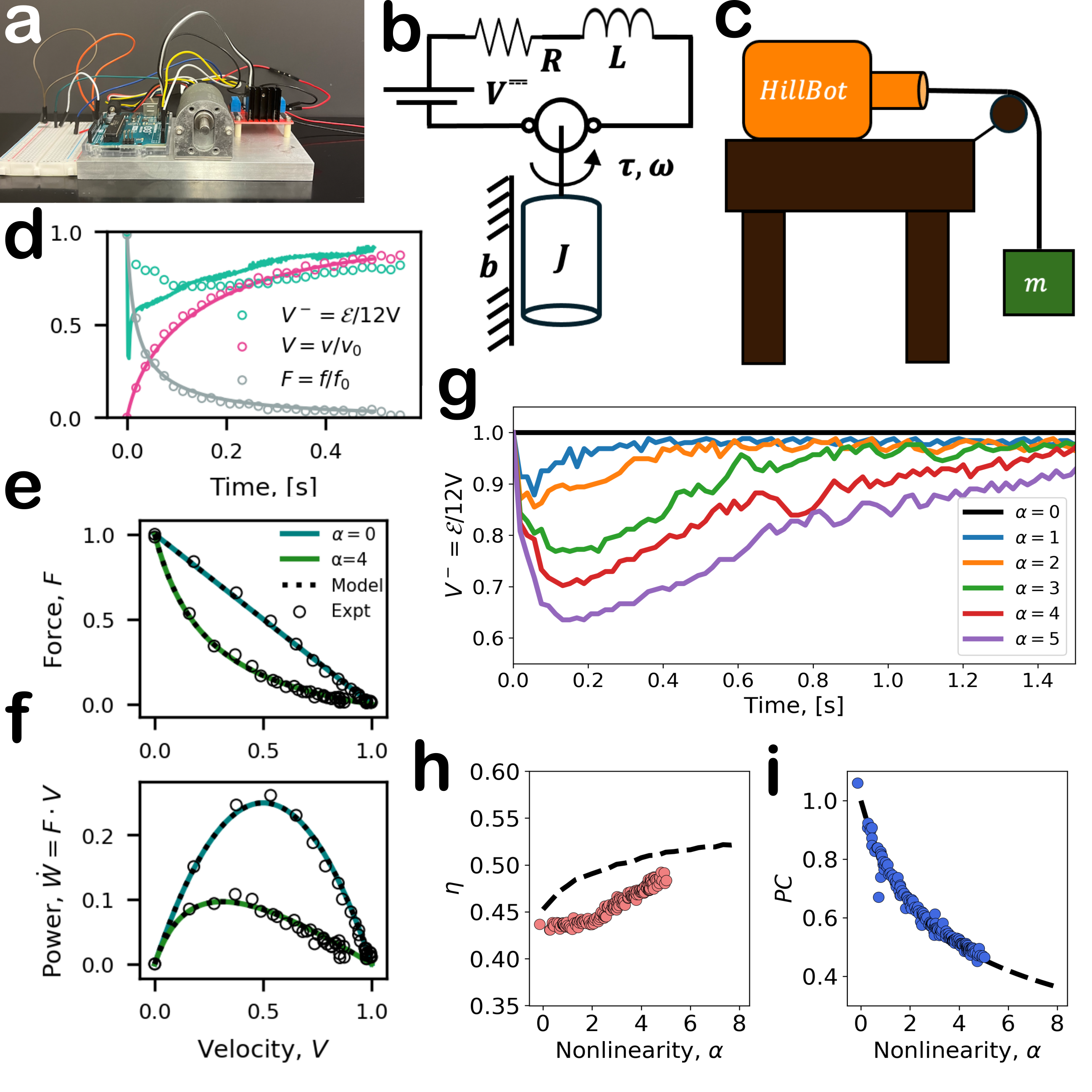}
    \caption{Introducing nonlinearity in HillBot demonstrates output and efficiency tradeoff. a) Photograph of HillBot. b) Schematic of electromechanical model of HillBot. c) Schematic of HillBot lifting a weight against gravity. d) Time series of data collected from HillBot; normalized, measured force (gray open circles), velocity (magenta open circles), and voltage (teal open circles) as a function of time overlayed on the model’s predictions (solid lines). e) $FV$ and f) $\dot{W}V$ curves for $\alpha=0,4$ given by Hill’s equation (blue and green lines, respectively), experimental measurements with HillBot (open circles), and electromechanical model (black dashed lines). g) The PID controller dynamically reduces voltage for increasing $\alpha$ to match Hill-type outputs (similar to cross-bridge unbinding in the 2-state model). h), i) HillBot's electromechanical model (black dashed lines) and $N=169$ trials in HillBot (scatter points) show an exchange of power characteristic (blue circles) and efficiency (red circles) with $\alpha$.}
    \label{fig:f3}
\end{figure}

So far we have demonstrated that increasing $\alpha$ simultaneously decreases power but increases efficiency. But because of these simultaneous effects, the mechanism underlying efficiency gains still remains unclear. Efficiency gains could result from the fact that lower-power actuators simply draw less energy. Alternatively, efficiency gains could result directly from nonlinearity. In order to disentangle $\alpha$’s simultaneous effects on power and efficiency, we perform a corresponding control experiment with an equivalent $PC$ actuated by a linear force-velocity profile (Fig. 4a, 4b), called "Series-L". These control experiments allow us to experimentally isolate the effect of the nonlinearity and disentangle $\alpha$’s concurrent effects on efficiency and performance. Furthermore, we conduct Series-L as a reference set of data to ensure that any findings through HillBot are the result of muscle’s inherent nonlinear FV-shape and not due to any unknown artifacts of our model system or due to the reduction in the actuator’s $PC$.

In the Series-L, we do not perform full-state feedback and actuate with a constant, reduced voltage, thus preserving the inherent linear $\alpha=0$ FV-relationship of the DC motor:
\begin{equation}
    F=C-V
\end{equation}
where the motor's normalized force and velocity $F=f/f_m, V=v/v_m \in [0,C]$ for $0 \leq C \leq 1$. To create Series-L, we reduce the constant input voltage $\mathcal{E}$ which effectively reduces $f_m$ and $v_m$ to a constant $C$ which we call the motor handicap (Fig. 4c) and thus reduces the motor’s power characteristic (Fig. 4d).

For each experiment with HillBot (with a given value $\alpha$), we perform a corresponding experiment in Series-L (with a value $C(\alpha)$) to produce a series of nonlinear-linear pairs that each have an equivalent $PC$ (Fig. 4a, 4b). A calculation shows $C = \sqrt{2} \cdot \sqrt{\ln(1+\alpha) \cdot (1+\alpha) - \alpha} / \alpha$ for any nonnegative $\alpha$ value (see SI for derivation).

Through Series-L, we find decreasing efficiency $\eta$ for decreasing $PC$ values (Fig. 4e, blue points), whereas in HillBot, efficiency increases as the $PC$ is reduced through the nonlinearity $\alpha$ (Fig. 4e, orange points). This result disentangles the effect of $PC$ on efficiency and suggests that any efficiency gains in HillBot are emergent from the curvature of Hill's force-velocity relationship and the dynamic "unbinding" of voltage.
\begin{figure}[H]
    \centering
    \includegraphics[width=0.6\textwidth]{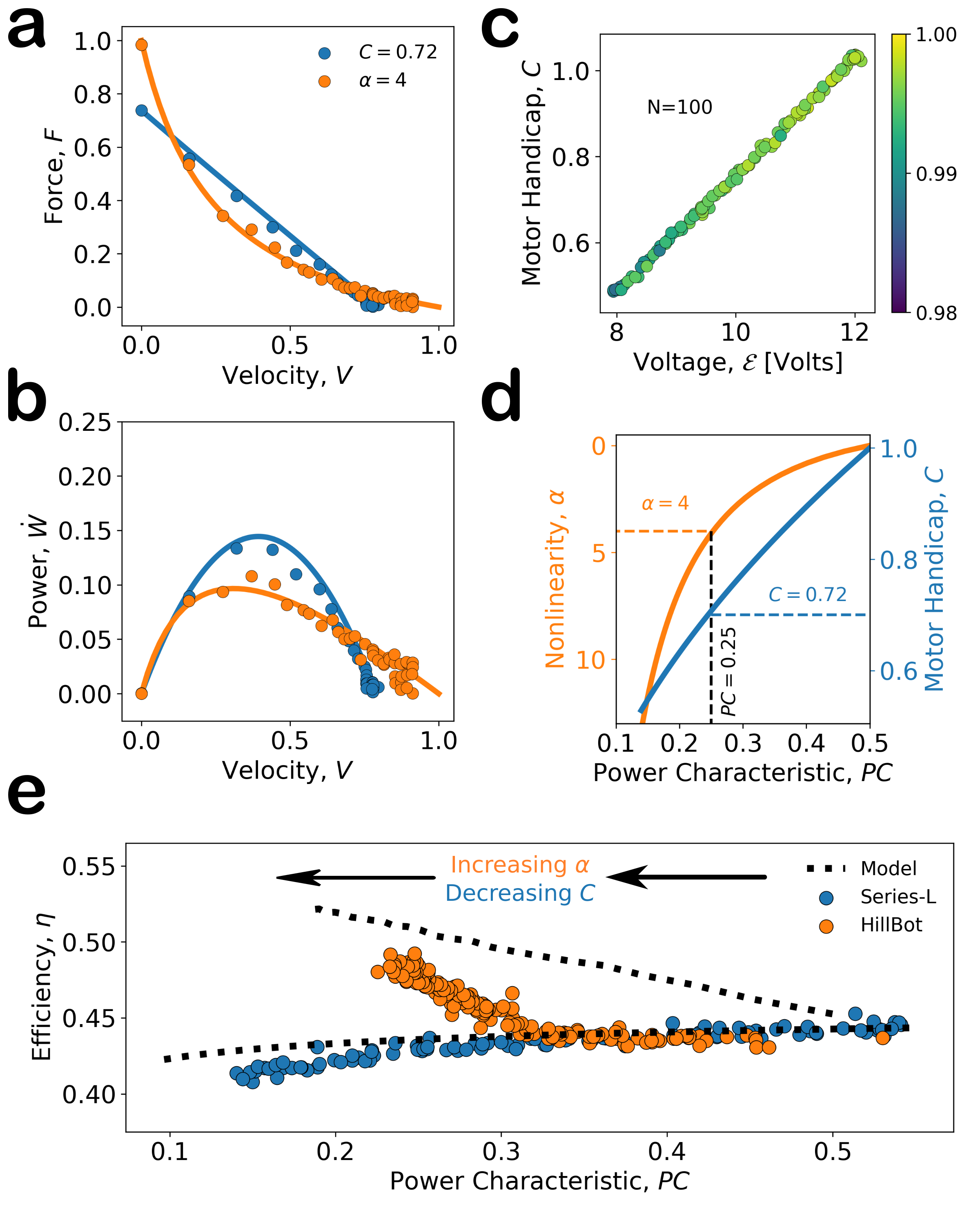}
    \caption{Series-L results. a), b) Comparison of linear $C=0.72$ and nonlinear $\alpha=4$ force-power-velocity relations for $PC=0.25$. c) Linear relationship between applied voltage and motor handicap; experimentally, we decrease voltage supply to decrease the motor constant to lower the $PC$ in Series-L. d) Motor constant $C$ and nonlinearity $\alpha$ values that generate FV-relations with an equal $PC$; for example, $PC(C=0.72)=PC(\alpha=4)=0.25$ as in 4a, 4b. e) Efficiency decreases with $PC$ in Series-L whereas efficiency increases with nonlinear actuation, disentangling $PC$'s effect on $\eta$. }
    \label{fig:f4}
\end{figure}
First, we started with a simple 2-state model of actin-myosin dynamics, and we were able to show that the molecular mechanism generating Hill's nonlinear force-velocity relation is the velocity dependent unbinding of myosin. We found that this nonlinearity sensitively parametrizes a tradeoff in efficiency ($\eta$) and performance ($PC$); low $\alpha$-valued muscles are powerful yet inefficient while high $\alpha$-valued muscles are efficient yet weak. Next, using a robophysical model of muscle, we reproduced these results in experiment and, moreover, through Series-L, showed that the efficiency gains of decreasing $PC$ are the result of muscle's nonlinear FV-shape. This result leads us to ask: what is the physiological significance of a tradeoff in efficiency ($\eta$) and performance ($PC$) in muscle and non-muscle cells? Is there any utility for it? How does this tradeoff, parameterized by the unbinding rate $\alpha$, manifest itself in nature?
\subsection{Observation of a characteristic nonlinearity $\alpha^*$}
Researchers have long known that muscle has a nonlinear force-velocity curve \cite{hill_heat_1938} (Fig. 5a, 5b). What values of the nonlinearity does naturally occurring muscle express? Hill initially speculated that the nonlinearity parameter $\alpha$ assumed a constant value of $\alpha \approx 4$ across all muscle. Since then, a broad range of $\alpha$-valued muscles have been reported in the literature. To understand how $\alpha$ is distributed across organisms, we perform a meta-analysis across $N=135$ individual measurements of $\alpha$ in biological muscle samples from 39 different studies \cite{de_ruiter_measurement_1999, gilbert_tension_1978, rall_energetics_1973, valour_influence_2003, miller_summary_2016, baratta_forcevelocity_1995, barclay_mechanical_1996, barclay_energetics_1993, houdijk_evaluation_2006, ranatunga_correlation_1990, ranatunga_force-velocity_1984, wilkie_relation_1949, hill_dynamic_1997, buschman_mechanical_1997, lannergren_forcevelocity_1978, lannergren_contractile_1987, mutungi_effects_1987, curtin_efficiency_1991, rome_influence_1990, nelson_forcevelocity_2004, cooke_inhibition_1988, woledge_energetics_1968, alcazar_shape_2019, gilbert_effect_1986, hill_heat_1938, hill_efficiency_1964, josephson_contractile_1987, asmussen_maximal_1989, caiozzo_influence_1991, claflin_force-velocity_1989, luff_dynamic_1985, tihanyi_force-velocity-power_1982, johnston_power_1984, josephson_contraction_1984, malamud_effects_1988, rome_influence_1992, romero_comparison_2016, winters_improved_1995}. (Fig. 5c, 5d, 5e, see Methods, SI Table 2). Across evolutionarily distant species, we observe that most muscle groups in our meta-analysis exhibit values of $\alpha$ near a characteristic value of approximately four ($\alpha^*=3.85 \pm 2.32$, median $\pm$ IQR). In addition to muscle tissue, Hill’s hyperbolic force-velocity relation has also been observed in individual mouse myoblast cells \cite{mitrossilis_single-cell_2009} where AFM experiments measured a value of $\alpha \approx 4$ (Fig. 5e, red dashed vertical line). Our meta-analysis aligns with Hill's original estimation of $\alpha \approx 4$, a value widely accepted in the literature \cite{mcmahon_muscles_1984}; however, our analysis also reveals a broader distribution of reported $\alpha$ values ranging from near zero to as high as 15. In order to classify muscle groups, we identify four distinct categories: muscle groups with a broad distribution of $\alpha$ (Fig. 5c, yellow), with a narrow distribution centered around the characteristic value $\alpha \approx 4$ (green), with lower values of $\alpha$ (blue), and with higher values (red). For the muscle groups that are present in humans, we depict their approximate locations in the human body (Fig. 5d). Although relating $\alpha$ to muscle function is not within the scope of this work, we discuss how nonlinearity and energetics could affect behavior on organismal scales (see Discussion).

\begin{figure}[H]
    \centering
    \includegraphics[width=0.75\textwidth]{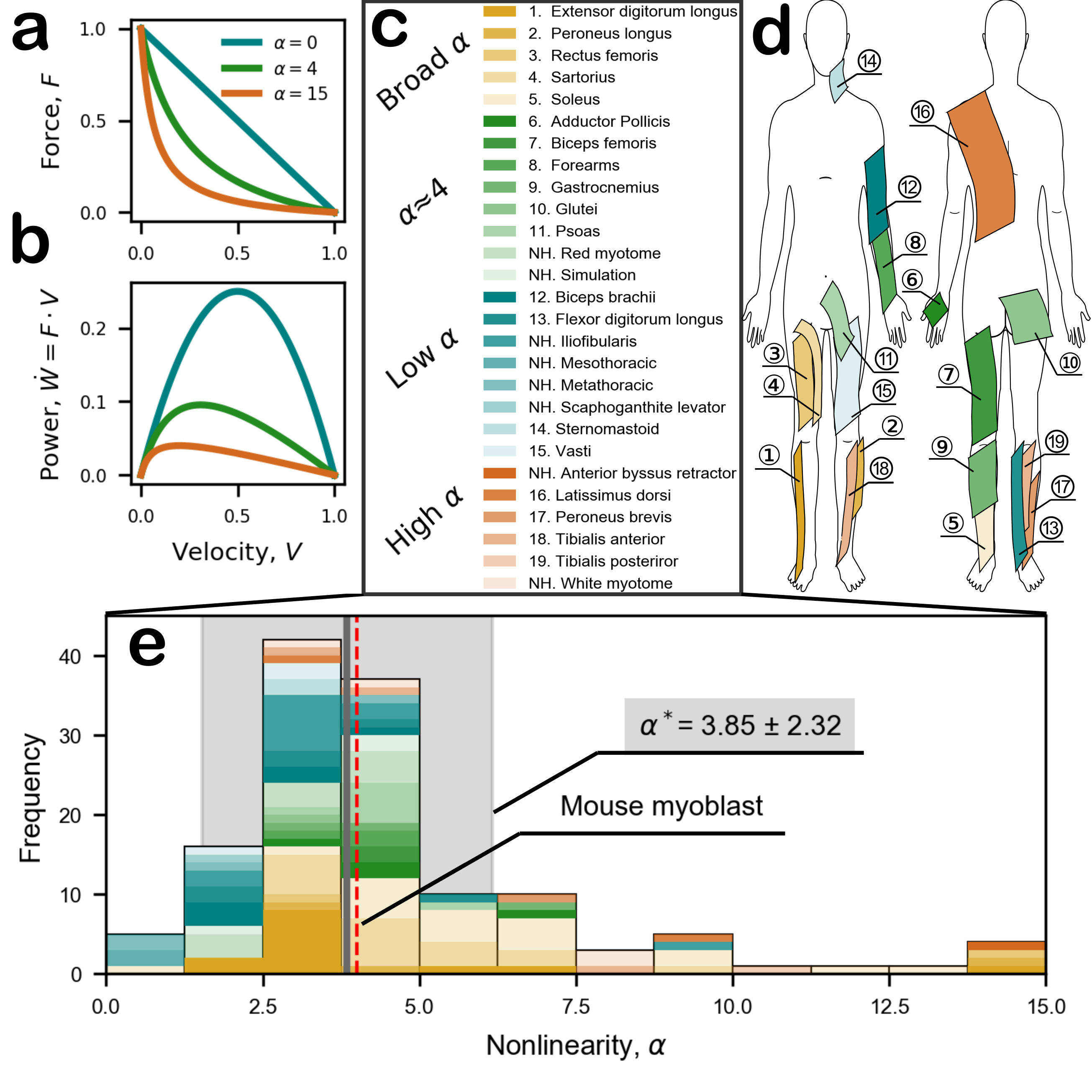}
    \caption{$\alpha$ meta-analysis results. a) Hill’s force-velocity ($FV$) and b) mechanical power-velocity ($\dot{W}V$) relations for $\alpha=0,4,15$. c) List of muscle types found in the meta-analysis. Muscles are split into four groups: those where $\alpha$ has a broad range (yellow), low values (blue), high values (red), and near $\alpha^*$ (green). Numbered label corresponds to d; those labeled NH indicate non-human muscles. d) Muscles found in the human body from the meta-analysis. Muscles are colored according to their label in c. e) Distribution of $\alpha$ in measured biological muscle tissue for $N=135$ muscle samples, including those from mammals, birds, mollusks, fish, amphibians, insects, and reptiles. Red dashed line indicates a measurement of the FV-relation from a myoblast cell. The nonlinearity $\alpha$ ranges between 0 and 15, centered at $\alpha^*=3.85 \pm 2.32$ (median, gray line $\pm$ IQR, shaded gray).}
    \label{fig:f5}
\end{figure}

The observation of a characteristic value of $\alpha^*=3.85 \pm 2.32$ (which notably excludes linear actuation $\alpha=0$) in muscle and myoblast cells raises the question whether this particular value confers energetic advantages. In order to provide a potential basis to explain this prevalence, we define a simple objective function $\phi_\text{Hill}(\alpha)$ that quantifies the performance-efficiency tradeoff in these actuators:
\begin{equation}
    \label{eq:e3}
    \phi_\text{Hill}(\alpha) := \left| \eta^\dagger(\alpha) - PC^\dagger(\alpha) \right|
\end{equation}
where $\eta^\dagger(\alpha)$ and $PC^\dagger(\alpha)$ are the normalized energetic efficiency and normalized power characteristic as a function of the nonlinearity $\alpha$. Energetic efficiency, by definition, is reported as a normalized quantity, $\eta^\dagger:=\eta/1$, and the power characteristic is normalized as $PC^\dagger:=PC(\alpha)/PC(\alpha=0)$. These normalization conditions ensure that maximally efficient actuators correspond to $\eta^\dagger=1$ and maximally powerful actuators correspond to $PC^\dagger=1$. The minimum of $\phi_\text{Hill}(\alpha)$ represents nonlinearities in FV-space that equally favor the power characteristic and efficiency.
\begin{figure}[H]
    \centering
    \includegraphics[width=0.75\textwidth]{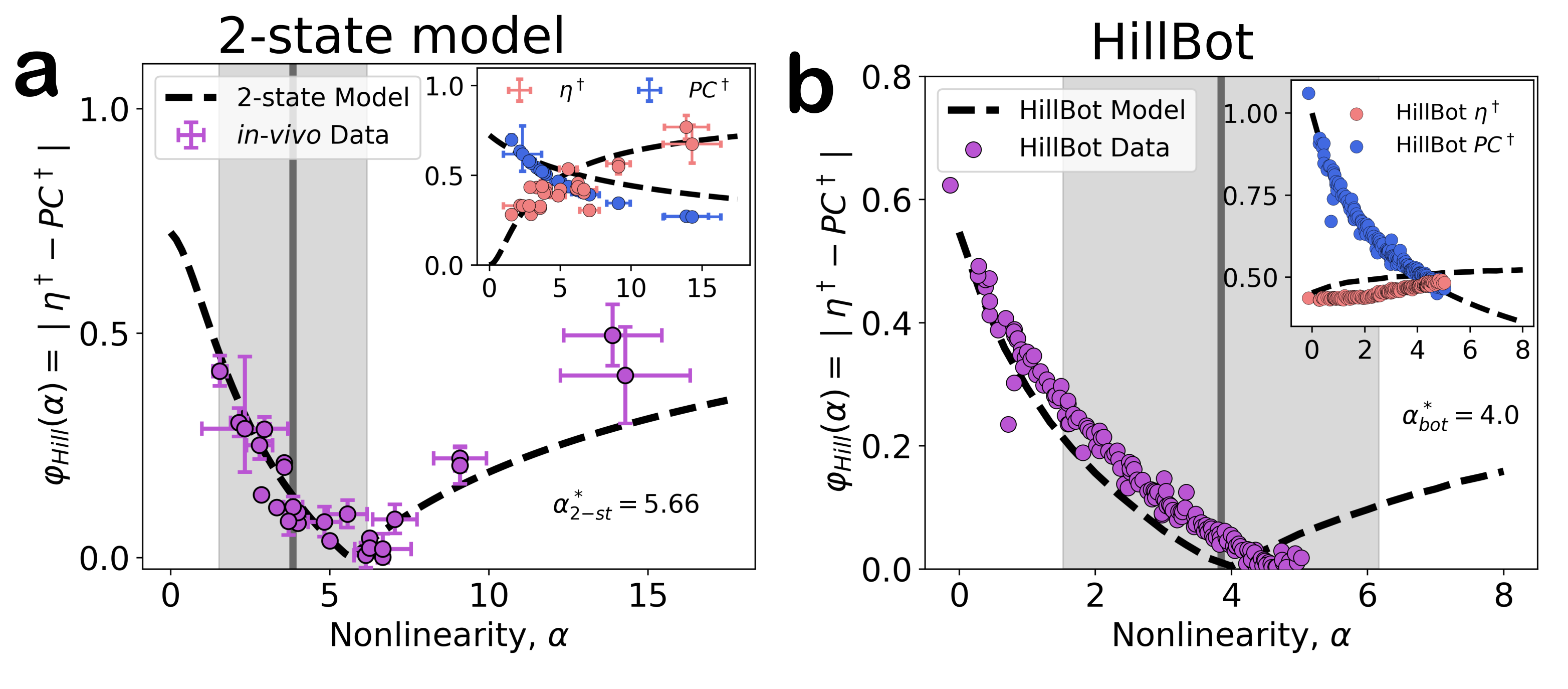}
    \caption{The performance-efficiency tradeoff quantified. a) Objective function applied to \textit{in-vivo} meta-analysis data (purple scatter points) and the simple 2-state actomyosin model (dashed black line). $\phi_\text{Hill}$ is minimized at $\alpha^*_\text{2-st}=5.66$, agreeing with $\alpha^*=3.85\pm2.32$ (gray shaded region). Inset: Efficiency $\eta$ (red symbols) and power characteristic $PC$ (blue symbols) as functions of the nonlinearity parameter $\alpha$. b) Objective function $\phi_\text{Hill}$ as a function of the nonlinearity parameter $\alpha$ for $N=169$ trials in HillBot (purple circles) with electromechanical model (black dashed line). Gray region corresponds to the characteristic nonlinearity $\alpha^*$. Sub-panel shows an exchange of power characteristic (blue circles) and efficiency (red circles) with $\alpha$; $\phi_\text{Hill}$ is minimized at $\alpha^*_\text{bot}=4.0$.}
    \label{fig:f6}
\end{figure}
We compute $\phi_\text{Hill}(\alpha)$ based on the $\eta$ and $PC$ data from our 2-state model and find a functional form that, by visual inspection, agrees with the results of the \textit{in-vivo} efficiency data from the literature (Fig. 6a). Our model predicts a minimum of $\phi_\text{Hill}$ at $\alpha^*_\text{2-st}=5.66$, agreeing with the characteristic value $\alpha^*=3.85 \pm 2.32$ from the meta-analysis. We further compute $\phi_\text{Hill}(\alpha)$ using the recorded $\eta$ and $PC$ data from our experiments with HillBot. We again observe a non-monotonic dependence of $\phi_\text{Hill}$ on $\alpha$, with a minimum at $\alpha^*_\text{bot}=4.0$ (Fig. 6b). This result again coincides with the characteristic nonlinearity $\alpha^*=3.85 \pm 2.32$ from our meta-analysis (Fig. 5e, gray-shaded region). These results raise the hypothesis that the most commonly observed nonlinearity $\alpha^* \approx 4$ balances performance and efficiency in muscle and contractile cells. However, testing this hypothesis in muscle cells would be difficult as it requires simultaneous measurement of energy consumption and mechanical power, as well as systematic tuning of $\alpha$ (or, respectively, myosin’s velocity-dependent unbinding rate) while keeping all other parameters constant.

\section{Discussion}
\subsection{Relating back to muscle}
Our results have investigated connections between myosin’s velocity-dependent unbinding dynamics and energetics (free energy consumption, mechanical power output, and efficiency). In our microscopic model, $\alpha = K_D/k_a$ controlled the nonlinear coupling between myosin unbinding dynamics and actuation of a coupled mass. Meanwhile, in HillBot, we implemented $\alpha$ by dynamically reducing voltage to mimic the concave force-velocity relation described by Hill. We demonstrated in our microscopic model that the nonlinearity parameter $\alpha = K_D/k_a$ describing myosin’s velocity-dependent unbinding rate resulted in transient dynamics resembling Hill’s concave force-velocity curve (Fig. 1b, 1c). Furthermore, in both experiments on HillBot (Fig. 3) and our microscopic model of muscle (Fig. 2), we found similar dependencies of mechanical power output $\dot W(\alpha)$ and efficiency $\eta(\alpha)$ on the nonlinearity parameter $\alpha$. Overall, our results demonstrate that nonlinear coupling between myosin binding dynamics and actuation impacts energetics, and that the energetics can be experimentally instantiated by mimicking Hill’s relation in robotic systems.

Although it is interesting that $\dot W(\alpha)$ and $\eta(\alpha)$ coincided between our microscopic nonlinear model of myosin and HillBot’s mimicking of muscle's concave force-velocity curve, our study does not attempt to relate myosin binding dynamics to Hill’s force-velocity curve for \textit{in vivo} systems. Although velocity-dependent unbinding does contribute to a concave force-velocity relation, other factors can affect the force-velocity curve. Coupling of muscle to the surrounding elastic connective tissue, and to elastic tendon, can affect the force-velocity curve \cite{hill_abrupt_1997}. Values of $\alpha$ can vary between individuals and species depending on fitness training \cite{andersen_changes_2005}, fatigue \cite{jones_changes_2010,kristensen_fatiguing_2019}, and temperature \cite{ranatunga_temperature_2018,ranatunga_force-velocity_1984,binkhorst_temperature_1977}. Muscle groups are recruited cooperatively rather than in isolation \cite{turpin_how_2021}. Furthermore, Hill’s original experiments were performed in the specific case where muscle contracts a constant load after relaxation from isometric tension \cite{hill_heat_1938}, and Hill's relation is not precise near stall \cite{edman_double-hyperbolic_1988, lou_high-force_1993}. A concave force-velocity curve can result from entirely different mechanisms as well, such as capillary forces \cite{cohen_capillary_2015}. Additionally, other muscle properties contribute to muscle energetics and function, \cite{zajac_muscle_1989} such as muscle's force-length relationship \cite{maganaris_forcelength_2001}, tendon's elastic energy storage \cite{abrahams_mechanical_1967}, coupling to multiple joints \cite{sreckovic_evaluation_2015,garcia-ramos_force-velocity_2016,feeney_loaded_2016}, Henneman’s size principle \cite{henneman_relation_1957}, and differences in metabolic supply of ATP \cite{mcmahon_muscles_1984}.

Relating microscopic-scale myosin binding dynamics to macroscopic-scale energetics and biological mechanical function would be an interesting future research direction. We encountered qualitative observations while performing our meta-analysis that could potentially lead to testable hypotheses. For muscle groups with broad distributions of $\alpha$ (Fig. 5c, yellow), and those with narrower distributions, close to the characteristic value of $\alpha \approx 4$ (green), we found some muscle groups that are implicated in a wide variety of tasks. The psoas (''11.''), for example, performs both dynamic and posture functions \cite{arbanas_fibre_2009}. Meanwhile, muscle groups with high values of $\alpha$ (Fig. 5c, red), such as the latissimus dorsi (''16.'') and the peroneus brevis (''17.’’), tend to sustain low levels of activity over long periods of time to assist in stabilization and posture regulation \cite{gerling_architectural_2013}, activities where energetic efficiency may be vital. Finally, muscle groups such as the biceps brachii (''12.'') and vasti (''15.''), exhibit almost linear force-velocity curves (Fig. 5c, blue). These muscle groups are mostly implicated in explosive, ballistic motions \cite{newton_kinematics_1996} where power generation is key. Again, we caution that these observations remain qualitative. However, we anticipate that studying nonlinear unbinding dynamics and energetics across muscle groups and species could yield interesting insights into how animal-scale function is sustained by regulation of molecular-scale energetic conversion. Furthermore, it would be interesting to compare and contrast muscle groups with low and high $\alpha$, and investigate whether their microscopic unbinding dynamics differ from the more common $\alpha \approx 4$ muscle groups.
\subsection{Variable $\alpha$ for improved control over energetic flows in robots}
Interestingly, our work shows that the power-efficiency tradeoff due to the nonlinearity $\alpha$ is not necessarily specific to actomyosin systems. In actomyosin, an increased fraction of myosin unbinds at higher contraction rates; similarly, HillBot reduces the applied voltage as the coupled mass increases velocity. Both responses amount to a “letting go” of actuation effort during transients. Although our analytical model of actomyosin dynamics included an expression for power consumption and heat generation, HillBot consumes and dissipates energy via entirely different mechanisms. Yet, we still found a performance-efficiency tradeoff in both systems. This result highlights the generality of nonlinear actuation’s control over energetic flows. We therefore propose to leverage HillBot’s actuation mechanism as a method of dynamically regulating energetic flows in robotics applications.

Seminal work demonstrated that dynamically regulating an actuator’s end-point impedance to mimic a linear damped-spring system stabilizes contact interactions \cite{hogan_impedance_1985}, and more recent work has highlighted the benefit of biologically-inspired hysteric damping \cite{brissonneau_biologically-inspired_2021,he_complex_2020}. Here we propose enhancing these approaches by incorporating dynamic control over the nonlinearity parameter $\alpha$ in variable impedance protocols. We anticipate that mimicking the nonlinearity affords high-level control over the energetic flows between an actuator and its coupled environment, dynamically biasing actuator-environment interactions along a performance-efficiency axis depending on context, energetic demands and reserves. This method improves existing designs which mimic a static value of $\alpha$ using hardware components \cite{pratt_series_1995,cohen_capillary_2015}.

One further consideration in impedance protocols relates to human safety and comfort in human-robot interactions, which are crucial factors in designing assistive devices \cite{gavette_advances_2024}. Human prosthetics need to carefully regulate power to avoid injury \cite{kikuuwe_guideline_2008}. Surprisingly, reducing prosthetic power does not impact the user's metabolic cost \cite{quesada_increasing_2016}. Moreover, amplifying the prosthetic's energetic efficiency allows for a lightweight, compact design \cite{alluhydan_functionality_2023}. Under a nonlinear actuation scheme with $\alpha$, the concavity of Hill's FV-relation simultaneously boosts efficiency and reduces power output (cf. Fig. 3h, 3i), addressing design and safety concerns without raising the user's metabolic cost. We anticipate that a variable nonlinear impedance control will result in more precise control over energetic flows, ultimately improving user experience with actuated prosthetics.

\section{Conclusions}
Here we have investigated the effect of the nonlinearity parameter $\alpha$, which governs myosin's velocity-dependent unbinding rate, on energetic flows during actuation. We demonstrated this quantification with a simplified theoretical model coupling myosin binding dynamics to a mass. The energetic predictions of this model agreed with measurements of 26 muscle samples from the literature. Moreover, the model illustrated a performance-efficiency tradeoff parameterized by the unbinding rate $\alpha$. Furthermore, we sought to mimic this velocity-dependent unbinding with HillBot, a robophysical model of Hill’s hyperbolic force-velocity curve. Experiments with HillBot allowed us to systematically control $\alpha$ while holding all other muscle properties constant. As in our actomyosin model, we found similar dependencies of the mechanical power output $\dot W(\alpha)$ and the efficiency $\eta(\alpha)$ on the nonlinearity parameter $\alpha$ in HillBot. Although the precise microscopic mechanisms by which $\alpha$ regulates actuation in both systems differ greatly, our main result appears to be generalizable: a negative feedback from velocity onto actuation (i.e. “letting go” or "unbinding") reduces power generation but boosts efficiency. Lastly, we compile 136 published measurements of $\alpha$ in muscle and myoblasts to reveal a distribution centered around $\alpha^* = 3.85 \pm 2.32$. Synthesizing data from our model and HillBot, we quantitatively show that the performance-efficiency tradeoff may underpin the prevalence of $\alpha^*$ in nature. We propose to leverage this understanding in variable-impedance protocols to enable real-time control over energetic flows in robotics applications. This high level of abstraction could provide more nuanced control over robot-environment and human-robot interactions, including prosthetics and exoskeletons.

\section{Methods}
\subsection*{Meta-analysis}
To acquire the meta-analysis data, we searched for peer-reviewed research articles that reported a measured value of $\alpha$ from a biological muscle tissue sample. The literature often describes force-velocity curvature as a ratio of the muscle’s coefficient of shortening heat to the muscle’s isometric force, using the notation $a/P_0$ or $a/F_0$ where $\alpha$ is the inverse of this ratio. For each study where FV-curvature is measured, we record the animal’s name and nickname, the muscle sample’s group (EDL, bicep brachii, etc.), the temperature at measurement (if available), and whether the measured $\alpha$ is the result of an average. Once we acquired all $N=135$ measurements of $\alpha$ in muscle (in addition to the lone $\alpha$ measurement in mouse myoblasts \cite{mitrossilis_single-cell_2009}), we created Fig. 5 to show the distribution of $\alpha$ in nature. Data from the meta-analysis is shown in SI Table 2. Furthermore, to generate Fig. 2 and Fig. 6, we analyze a subset of $N=26$ samples where energetic efficiency and FV-curvature are reported for the same muscle sample (see SI Table 4).
\subsection*{HillBot}
HillBot is a physical model of the Hill muscle that uses feedback control to mimic muscle's nonlinear force-velocity behavior in Eq. 1. Fig. 3 depicts HillBot and SI Fig. 1 shows the control scheme implemented for biomimetic actuation. Our methods in physical-model development and data acquisition follow from our previous paper \cite{mcgrath_hill-type_2022}.

Hardware: HillBot comprises a motor (Pololu 70:1 gear motor) with rotary encoder (4480 counts per revolution resolution). A time series of positional changes are used to calculate the motor’s velocity. The measured velocity is analogous to muscle contraction rate. A Hall effect-based current sensor (ACS714 Current Sensor $\pm$5 A) with low resistance ($\sim$1.2 m$\Omega$) is placed in series with the motor and reads the motor’s electrical current. The torque applied by a DC motor is proportional to the current and is analogous to muscle force.

Software: We program a microcontroller (Arduino UNO R3) with proportional-integral-derivative (PID) control with optimized gain coefficients to mimic muscle's nonlinear FV-behavior (SI Fig. 1a). At each time step (loop sampling rate $\sim$10 ms), we record force and velocity using the current sensor and encoder. The controller maps system state (force and velocity) to mimic Hill's relation for a given initialized $\alpha$ value by dynamically controlling the applied voltage $\mathcal{E}$.

Normalized state variables: For congruence with the normalized Hill model in Eq. 1, we normalize measured force and velocity by $V=v/v_m$ where $v$ and $v_m$ are the measured and maximum angular velocities of the motor. To normalize the measured force, we use $F=(i-i_s)/(i_m-i_s)$ so that force $F$ is proportional to current $i$ as expected. We subtract off the steady state current $i_s$ (that is, $i$ when the motor is at $v_m$) so that the actuator produces little force close to the no-load velocity.

Model: We start with Newton’s and Kirchhoff’s second laws to construct differential equations describing the electo-mechanical behavior of our actuator (Eq. 10-13). The PID gains ($k_p,k_i,k_d$) are determined through trial and error until the model produces the expected Hill-type dynamics. The resistance, inductance, angular velocity-back EMF constant, torque-current constant, moment of inertia, and damping constants (written as $\Pi={R,L,k_1,k_2,J,b}$) are unknowns. We determine these constants by supplying a constant voltage to the system and recording system state. We sweep through a range of voltages from 12 to 8 Volts in $N=100$ trials. For each set of time series data, we perform a nonlinear fit using LMFIT (a nonlinear least-squares minimization Python package) to determine the six parameters $\Pi$ (SI Fig. 1c).

Finally, we compare the characteristics of our experimental data (scatter points) and model (black dashed line) with the predictions of Hill's equation for different $\alpha$ values (SI Fig 1b). For a given value of $\alpha$, we record a time series of the actuator’s force-velocity data and mechanical power-velocity data. Both curves are integrated, and we extract the maximum power output of the actuator for each $\alpha$. We then compare how the area under the actuator’s $FV$ and $\dot{W}V$ curves and maximum power output agree with Hill’s original muscle model in Eq. 1. The results in SI Fig 1b show strong agreement between our model, data, and Hill’s muscle model for all tested values of $\alpha$, demonstrating that our system accurately mimics the mechanics of Hill's model.

\newpage
\bibliographystyle{unsrt}
\bibliography{refs}

\subsection*{Author Affiliations}
Department of Physics, University of Texas at Austin, 2515 Speedway, Austin, Texas 78712, USA
\subsection*{Acknowledgments}
We would like to thank William Gilpin, Richard Neptune, Owen Beck, Walter Herzog, Melissa Kemp, and Luis Sentis for helpful discussions. We would also like to thank Chris McGrath for helping design Fig 5d. This material is based upon work supported by the National Science Foundation under Grant No. DMR-2144380, NSF PHY-2309135, and the Gordon and Betty Moore Foundation Grant No. 2919.02 to the Kavli Institute for Theoretical Physics (KITP).
\subsection*{Author Contributions}
JM and JA designed experiments. JM and BK preformed research; JM analyzed data. JM, CJ, and JA developed models. JM and JA wrote manuscript.
\subsection*{Competing Interests Statement}
Authors declare no competing interests.
\end{document}